\documentclass[aps,pre,twocolumn,superscriptaddress]{revtex4}
\usepackage{setspace}
\usepackage{amsmath}
\usepackage{amssymb}
\usepackage{epsfig}
\begin{document}
\preprint{}
\title{Measuring the Nematic Order of Colloidal fd Virus by X-ray Diffraction}
\author{Kirstin R. Purdy}
\affiliation{Complex Fluids Group, Department of Physics, Brandeis
University, Waltham, Massachusetts 02454}
\author{Zvonimir Dogic}
\affiliation{Complex Fluids Group, Department of Physics, Brandeis
University, Waltham, Massachusetts
02454}\altaffiliation{Department of Physics, University of
Pennsylvania, Philadelphia, Pennsylvania}
\author{Seth Fraden}
\affiliation{Complex Fluids Group, Department of Physics, Brandeis
University, Waltham, Massachusetts 02454}
\author {Adrian R\"{u}hm}
\affiliation{Max-Planck-Institut f\"{u}r Metallforschung,
Heisenbergstrasse 1, D-70569, Stuttgart, Germany}
\author{Lawrence Lurio}\affiliation{Department of Physics, Northern Illinois University,
DeKalb, Illinois 60115 }
\author{Simon G. J. Mochrie}
\affiliation{Department of Physics, Yale University  New Haven,
Connecticut 06520}
\date{\today}
\begin{abstract}

The orientational distribution function of the nematic phase of
the semi-flexible rod-like virus fd is measured by x-ray
diffraction as a function of concentration and ionic strength.
X-ray diffraction from a single-domain nematic phase of fd is
influenced by interparticle correlations at low angle while only
intraparticle scatter contributes at high angle. Consequently, the
angular distribution of the scattered intensity arises from only
the single particle orientational distribution function at high
angle but it also includes spatial and orientational correlations
at low angle. Experimental measurements of the orientational
distribution function from both the interparticle and
intraparticle scattering were made to test whether the
correlations present in interparticle scatter influence the
measurement of the single particle orientational distribution
function. It was found that the two types of scatter yield
consistent values for the nematic order parameter. It was also
found that x-ray diffraction is insensitive to the orientational
distribution function's precise form, and the measured angular
intensity distribution is described equally well by both Onsager's
trial function and a Gaussian. At high ionic strength the order
parameter $S$ of the nematic phase coexisting with the isotropic
phase approaches theoretical predictions for long semi-flexible
rods $S=0.55$, but deviations from theory increase with decreasing
ionic strength. The concentration dependence of the nematic order
parameter was also found to better agree with theoretical
predictions at high ionic strength, indicating that electrostatic
interactions have a measurable effect on the nematic order
parameter. The measured x-ray order parameters are also shown to
be proportional to the measured birefringence and the saturation
birefringence of fd is measured, enabling a simple, inexpensive
way to measure the order parameter.
\end{abstract}
 \maketitle

\section{Introduction}
The first step to understand the effect of steric interactions in
colloidal rod systems was taken in 1949, when Onsager wrote his
seminal paper on the phase behavior of hard and charged rods
\cite{Onsager49}. Onsager developed a free energy theory at the
second virial level describing the phase transition of hard rods
from an isotropic phase, in which the particles are randomly
oriented, to a nematic phase, in which particles are oriented in a
distribution about a preferred direction. All theoretical
predictions for the properties of this phase transition, such as
the coexistence concentrations and the nematic order parameter,
depend on the functional form of the orientational distribution of
the rods in the nematic phase. Onsager chose one test function and
in a later review paper Odijk showed that qualitatively similar
results for the properties of the phase transition can be found by
choosing a Gaussian test function \cite{Odijk86}. The exact form
of the orientational distribution function that satisfies the
Onsager theory can be obtained via series expansion
\cite{Lasher70,Kayser78,Lekkerkerker84} or by direct iterative
methods \cite{Herzfeld84,lee86}. Determining the orientational
distribution function of the nematic phase of a colloidal rod
system is the most sensitive test of whether a system is described
by Onsager's theory.

In this paper we measure the measure the concentration and ionic
strength dependence of the orientational distribution function of
fd virus via x-ray diffraction. The fd virus is a charged
semi-flexible rod with a length $L$ to diameter $D$ ratio
$L/D\approx130$. The electrostatic charge on the rods can be taken
into account by an effective diameter
$D_{\mbox{\scriptsize{eff}}}$, larger than the bare diameter,
which is approximately equal to the distance between particles
when the interaction potential is about
$k_{\mbox{\scriptsize{B}}}T$. The exact calculation for the
effective diameter is outlined in Ref.
\cite{Onsager49,Tang95,Stroobants86a}. An increase in ionic
strength of the solution containing the charged rods produces a
decrease in effective diameter. In Onsager's theory, the limit of
stability of the isotropic phase is predicted to be $\pi/4
nD_{\mbox{\scriptsize{eff}}}L^{2}=4$, where $n$ is the number
density \cite{Kayser78}. This is predicted to be valid for long
rods with a length to effective diameter ratio greater than 100
\cite{Straley73}. Previously the isotropic and nematic coexistence
concentrations fd have been measured as a function of ionic
strength and it has been shown to agree well with numerical
results from Chen for a semi-flexible rod of with a ratio of
persistence to contour length of 2.5 \cite{Tang95}. Theoretical
models suggest that semi-flexibility also acts to significantly
lower the nematic order parameter at coexistence. For fd, a
relatively rigid polymer with a ratio of persistence to contour
length of 2.5, the nematic order parameter at coexistence is
predicted to be $S= 0.55$, which is significantly smaller than
predicted for rigid rods, approximately $S=0.79$ \cite{Chen94}.
Several review articles describe in more detail the theoretical
and experimental aspects of this and other systems described by
Onsager's theory
\cite{Straley73,Odijk86,Gelbart80,Lekkerkerker93r,Sato96,Fraden95,Forsyth78,vandermaarel95,Davidson97}.

In x-ray diffraction, the scattered intensity consists of two
parts, intraparticle scatter $F(\vec{q})$ and interparticle
scatter $S(\vec{q})$. The intensity can be written as a product of
the two types of scatter
\begin{eqnarray}\label{IequalsFS}
I(\vec{q})=NF(\vec{q})S(\vec{q},f(\vec{q}))
\end{eqnarray}
where $\vec{q}$ is the three dimensional reciprocal vector in
cylindrical coordinates $\vec{q}=(q_r, q_z, \phi)$. In a uniaxial
nematic, $q_r$ is perpendicular to the nematic director and the
scattered intensity is independent of the azimuthal angle $\phi$
about the director. If the system is oriented such that the
nematic director is in the $\hat{z}$ direction, $\vec{q}$ can be
described by $\vec{q}=(q_r,q_z)$. The intraparticle interference,
or form factor, contains information about the structure of the
individual particles. $F(\vec{q})$ can also be written as
$<f(\vec{q})^{2}>$ where $f(\vec{q})$ is the fourier transform of
the electron density of a particle and the average is over all the
particles and their orientations. The interparticle interference,
or structure factor term, contains information about the
positional and orientational correlations between particles. The
structure factor depends on the positions of the centers of
gravity of two scatterers $\vec{R_i}, \vec{R_j}$ and their
relative orientations \cite{Maier92}:
\begin{eqnarray}
S(\vec{q})=1+\frac{1}{N F(\vec{q})}<\sum_{i\neq j}^{N}
e^{i\vec{q}(\vec{R_i}-\vec{R_j})}f_{i}(\vec{q})f_{j}(\vec{q})>
\end{eqnarray}
The orientation of the particles is included in $f(\vec{q})$ and
the average $<>$ is over all particles and their orientations. For
isotropic scatterers, $f_i(\vec{q})=f_j(\vec{q})$ and the
structure factor and the form factor decouple, but for anisotropic
scatterers, $f_i(\vec{q})\neq f_j(\vec{q})$ unless the particle
orientations are the same. Therefore, in contrast to scatter from
spheres, the structure factor $S(\vec{q})$ of rods can not, in
general, be decoupled from its anisotropic form factor
$F(\vec{q})$.

In a nematic system, however, there is no long ranged
translational order. As a result, $S(\vec{q})$ approaches unity in
the limit of high $\vec{q}$, and if $S(\vec{q})=1$, the scattered
intensity is due only to the intraparticle interference
diffraction and $I(\vec{q})=F(\vec{q})$. In this regime the
angular distribution of the scattered intensity is a function only
of the single particle orientational distribution function.
Because of the crystalline internal structure of viruses such as
fd and Tobacco Mosaic Virus (TMV), x-ray diffraction produces a
complex pattern of intraparticle scatter at high $\vec{q}$ which
can be used to measure the single particle orientational
distribution function of the viruses \cite{Oldenbourg88}.

At low $\vec{q}$ the scattered intensity is dominated by
$S(\vec{q})$, and the angular distribution of the interparticle
interference scatter is influenced by the angular and spatial
correlations between neighboring rods. When intraparticle
interference scattering is absent or too weak to interpret, as in
thermotropic liquid crystal systems \cite{Leadbetter78}, or the
system of lyotropic vanadium pentoxide ($\mbox{V}_2\mbox{O}_5$)
\cite{Davidson97}, x-ray investigations of the nematic
orientational distribution rely on measuring the angular
distribution from interparticle interference scattering. In this
case one does not calculate the single particle orientational
distribution function, but instead the coupled fluctuations of
neighboring rods; this is predicted to overestimate the value of
the nematic order parameter for highly ordered samples
\cite{Leadbetter78, Davidson95}.

In this paper we explore the behavior of the nematic phase of fd
virus, investigating the concentration and ionic strength
dependance of the spatial and orientational ordering. We present
measurements of the orientational ordering of the nematic phase in
coexistence with the isotropic phase as a function of ionic
strength and compare the results with the predictions for
semi-flexible rods. Previously, measurements of the orientational
distribution function of a nematic phase have been made either
from form factor scatter as in work done by Oldenbourg et al on
TMV \cite{Oldenbourg86} and work done by Groot et al and
Kassapidou et al on persistence lengthed DNA fragments
\cite{vandermaarel94, vandermaarel95} or from structure factor
scatter as in work done by Davidson et al \cite{Davidson97}. Using
fd as our model rod allows us to not only measure the
orientational distribution function from intraparticle scattering,
but also from interparticle interference scattering. This allows
us to experimentally resolve this question of whether or not
correlations between angular and spatial order present in
interparticle scatter influence the measurement of the order
parameter. The saturation birefringence of fd was also measured
allowing for measurements of the order parameter to be made using
birefringence methods which involve much simpler techniques than
x-ray diffraction.

This paper is organized in the following manner. In section
\ref{mm} we describe the virus system and the experimental
methods. In section \ref{obs} qualitative observations about the
diffraction data are made. This is followed by a description of
the analysis technique used to extract the orientational
distribution function from the diffraction data in section
\ref{analy}. Quantitative measurements of the nematic spatial
ordering and orientational ordering are presented in section
\ref{results}. This includes first a brief section describing the
measured spatial ordering and then a section presenting the
measured orientational ordering of the nematic fd. Section
\ref{concl} summarizes the significant results of this paper.

\section{Materials and Methods \label{mm}}
The physical characteristics of the bacteriophage fd are its
length $L$= 880 nm, diameter $D$= 6.6 nm, persistence length $p$=
2200 nm and charge per unit length of around 10$\mbox{e}^{-}/$ nm
at pH 8.2 \cite{Fraden95}. When in solution, fd exhibits
isotropic, cholesteric, and smectic phases with increasing
concentration \cite{Lapointe73,Wen89,Dogic97,dogic00c}. The fd
virus was prepared using standard biological protocols found in
Ref. \cite{Sambrook89} using the JM101 strain of E. coli as the
host bacteria. The standard yield is $\approx$50 mg of fd per
liter of infected bacteria, and virus is typically grown in 10-12
liter batches. The virus was extensively dialyzed against a 20 mM
Tris-HCl buffer at pH 8.2 and the ionic strength was adjusted by
adding NaCl.

X-ray diffraction was done at the SAXS station on beamline 8-ID at
the Advanced Photon Source at Argonne National Lab. The beam flux
is $2\times10^{10}$ photons/s for a $50\times50 \mu$m beam with a
photon energy of 7.664 KeV ($\lambda$=1.617 \AA). The samples were
a suspension of monodisperse fd in the cholesteric phase, sealed
in $\approx0.7$ mm quartz x-ray capillaries. Cholesteric samples
were unwound and aligned in a 2 T permanent magnet (SAM-2
Hummingbird Instruments, Arlington, MA 02474)\cite{Oldenbourg86},
forming a single domain nematic phase parallel to the long axis of
the capillary and the magnetic field, which we will call
$\hat{z}$. The free energy difference between the cholesteric and
nematic phases is negligible, and the theory of the phase behavior
of the isotropic to nematic transition can be applied equally well
to the isotropic to cholesteric transition observed in fd
\cite{Tang95}. The magnetic field does not have a significant
effect on the ordering of the nematic phase
\cite{Torbet81,deGennes93,Tang93}. Samples had to remain in the
magnetic field for a minimum of 15 minutes at low concentrations
and a maximum of about 8 hours at the highest concentrations. The
strength of the magnetic field limited the maximum concentration
at which we could unwind the cholesteric phase into a mono-domain
nematic to about 100 mg/ml \cite{dogic00c}. Alignment of the
nematic sample was checked in a polarizing microscope and, using a
3$\lambda$ Berek compensator, its optical retardance was measured.
To easily view the solution within the capillaries under the
microscope, samples were placed in an index matching water bath
while still within the magnet. Birefringence is related to optical
retardance $R$ by $\Delta n = R/d$; $d$ is the measured thickness
of the nematic sample within the capillaries. The magnet and
sample were then mounted in a vacuum chamber such that the sample
was in the beam line, and the magnetic field was perpendicular to
the incoming beam. To observe the effect of charge on the nematic
phase, samples were prepared at different concentrations and ionic
strengths. The fd concentration was measured with a UV
spectrometer by absorption at 269 nm with an absorption
coefficient of 3.84 cm$^2$ mg$^{-1}$.

When the solutions of fd were exposed to x-rays for extended time,
disclination lines that matched the pattern traced by the beam
could be seen with a polarizing microscope. Since our samples were
exposed for varied times, a series of x-ray diffraction patterns
from the samples were collected with increasing x-ray exposure
time to quantify sample damage and its effects on the scattering
pattern. The polarizing microscope revealed sample changes after
$\approx6$ s of exposure, but the angular spread of the
diffraction peaks was not affected until exposure times increased
above 10 s, at which point the angular interference peak scatter
broadened significantly. The effect of exposure for $<$ 10 s on
the calculation of the order parameter was not measurable. Data
was collected for the interparticle interference scatter by
averaging ten 10 s exposures taken at different $50\times50 \mu$m
sections. To observe the much less intense intraparticle scatter,
the sample was continuously moved through the $50\times50 \mu$m
beam allowing for a total exposure of 120 s. A single long
exposure was used to image intraparticle scatter because it
resulted in less noise than multiple short exposures because
readout noise on the CCD was higher than the dark current. Readout
noise and solvent scatter were subtracted from data images during
analysis, but over the q-range which was analyzed this background
scatter was very uniform and could be approximated as a constant.

\section{Observations \label{obs}}
\begin{figure}
\epsfig{file=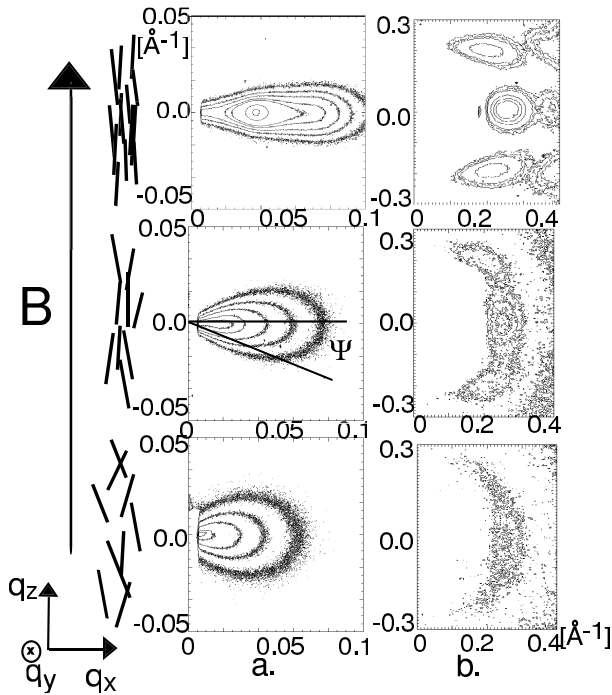,width=8cm}
\caption[]{\label{sfbw1.fig} (a) Contour plots of scattering from
nematic fd due to interparticle scatter.(b) Contour plots of
scattering from nematic fd due to intraparticle interference. The
interparticle scatter shown in (a) is hidden behind the beamstop
in (b) which is located on the left of the images and can not be
seen in the contour plot. From top to bottom the concentration of
the samples are 93 mg/ml, 33 mg/ml, and 15.5 mg/ml. Samples shown
are at an ionic strength of 10 mM (20mM Tris buffer), pH 8.2. The
magnetic field and virus orientation are perpendicular to the
scatter as shown in the schematic. $\Psi$ is the angle from the
equator on the detector film. Axes of contour plot are labeled in
\AA$^{-1}$, note the scales are different in (a) and (b).}
\end{figure}

The two dimensional scattered intensity of low angle interparticle
and high angle intraparticle interference peaks are shown in Fig.
\ref{sfbw1.fig} for concentrations spanning the range over which
fd is nematic at 10 mM ionic strength. The angular spread of both
types of scatter broadens with decreasing fd concentration or
increasing ionic strength, corresponding to an increase in
disorientation of the rods. The low angle peak, in Fig.
\ref{sfbw1.fig}a, is very intense and is a result of interparticle
interference. The horizontal position of this peak is inversely
proportional to the average interparticle separation, and the
radial width of the peak is inversely proportional to the
correlation length of the interacting rods. At larger scattering
angle, the fd layer lines are visible as shown in Fig.
\ref{sfbw1.fig}b. These intraparticle peaks are much less intense
than the interparticle interference peaks and are the result of
single particle scatter arising from the helical packing of the
virial proteins. The layer lines occur at intervals along the
$\hat{z}$ direction proportional to the reciprocal of the axial
repeat of the helical protein coat, which is 33\AA
\cite{Makowski81}. Because of discrepancy in both intensities and
scattering angle between the interparticle and intraparticle
scatter, we were unable to image both the high and low angle
scatter simultaneously.

Because of the short ranged positional order in the nematic phase
these intraparticle interference peaks should be independent of
interparticle correlations. We confirmed this hypothesis by
observing that the location of the peaks do not change with
concentration as the interparticle peaks do. We also compared our
data to published fiber diffraction results for
M13\cite{Glucksman92,makowski_note}. M13 is also a filamentous
bacteriophage, which only differs from fd by one amino acid per
coat protein: their structures are otherwise identical and
virtually indistinguishable by x-ray fiber diffraction
\cite{Makowski92}. Upon comparing published fiber diffraction data
with our data from nematic fd, we observed that they were similar,
but that the fiber diffraction patterns had Bragg peaks due to the
hexagonal packing of the virus in the fiber which were absent in
our nematic diffraction data. We also noticed that the horizontal
location of the single particle peaks in the fiber diffraction was
4\% larger than the location of our solution diffraction peaks,
indicating that the fiber diffraction was done on virus which had
a smaller diameter than those in our nematic samples. The fibers
are partially dehydrated, so it is not surprising that they become
compressed. The layer line spacing, however, was not altered,
indicating that no stretching of the virus occurs in the fibers.
From these observations we concluded that the high angle scatter
from the nematic fd was independent of interparticle correlations.
Detailed analysis of both the interparticle and intraparticle
diffraction continues in the following sections.

\section{Orientational Analysis Technique \label{analy}}
Because of the short ranged positional order of the nematic phase,
the high angle scattered intensity should be independent of
interparticle correlations $S(q_{r},q_z)=1$. We have demonstrated
above that this is true for fd. In this case, the intraparticle
scattered intensity of a system of rods is related to the
orientation of those rods in the following manner
\cite{Oldenbourg88, Holmes74}:
\begin{eqnarray}
\label{I_q} I(q_{r},q_z) =<I_s(q_r,q_z)>=\int
{\Phi}({\Omega}){I_s(q_{r}[\Omega],q_z[\Omega])} d\Omega
\end{eqnarray}
\noindent Where $\Omega$ is the solid angle ($\theta,\phi$) a rod
makes with respect to the nematic director $\theta$ and
azimuthally with respect to the incident beam $\phi$. Because fd
is axially symmetric $\Phi(\Omega)$ simplifies to $\Phi(\theta$).
${I_s(q_r,q_z)}$ is the axially symmetric three dimensional form
factor $f(\vec{q})^2$ of a single rod. $\Phi(\theta)$ is the
orientational distribution function (ODF) of the rods. Because the
form of the ODF is not known exactly, three test functions were
used:
\begin{eqnarray}
{\Phi}({\theta})=&A\exp{-\frac{{\theta}^{2}}{2{\alpha}^{2}}}&(0\leq \theta \leq \pi/2)\nonumber\\
=&A\exp{-\frac{(\pi-{\theta})^2}{2{\alpha}^2}}&(\pi/2 \leq \theta \leq \pi)\;, \\
{\Phi}({\theta})=&A\exp{-\frac{(\sin{\theta})^{2}}{2{\alpha}^{2}}}&(0\leq \theta \leq \pi) \;,\\
\label{onsODF}{\Phi}({\theta}) =& \frac{\alpha
\cosh{\alpha\cos{\theta}}}{4 \pi \sinh{\alpha}}&(0\leq \theta \leq
\pi)\;,
\end{eqnarray}
\noindent where $\alpha$ sets the width of each of the peaked
functions, and $A$ is the normalization constant such that $\int
{\Phi}(\theta)\sin(\theta)d\theta d\phi =1$. Eq. \ref{onsODF} is
normalized. The first ODF is the Gaussian used by Odijk
\cite{Odijk86}, the second is the function used by Oldenbourg et
al. \cite{Oldenbourg88} in their study of diffraction from nematic
TMV, and the third was defined by Onsager. The nematic order
parameter,
\begin{eqnarray}
S=2\pi\int_{0}^{\pi}\left(\frac{3}{2}\cos^{2}(\theta)
-\frac{1}{2}\right)\Phi(\theta)d\cos(\theta)
\end{eqnarray}
\noindent was determined for the orientational distribution
functions which best described the diffraction patterns.

The scatter from intraparticle interference was analyzed by
comparing it to a simulated scatter created from the evaluation of
Eq. \ref{I_q} using a three dimensional model for the single rod
form factor and a trial ODF. Previously Oldenbourg et. al.
measured the ODF from the intraparticle interference scatter of
TMV by simplifying Eq. \ref{I_q} to a one dimensional integral at
a constant $q_r$ \cite{Oldenbourg88}. This one dimensional method
could not be used for intraparticle fd scatter because fd has a
protein coat with a pitch  much larger than that of TMV, $33 \AA$
versus $23 \AA$ respectively, resulting in layer line overlap at
low concentrations. Instead, the radial intensity distribution of
single rod was modeled by \cite{Holmes74}
\begin{eqnarray}
I_s(q_r,q_z)=I_{m}(q_r,q_z)\sqrt{2\pi}\alpha q_{r}.
\end{eqnarray}
\noindent $I_m$ is the scattered intensities along the middle of
the zeroth and $\pm$ first layer lines of our most aligned nematic
sample, $S=0.96$ and Gaussian $\alpha=0.11$ as determined by the
interparticle interference peak. The intraparticle interference
data that fell on the detector in the range of $q_r=0.19 -0.33$
\AA, which encompasses the innermost peak on each of the three
layer lines visible in the interference pattern, was fitted to the
model diffraction images. For each diffraction pattern, an
$\alpha$ was found for each trial distribution function which
minimized a computed chi-squared value
\begin{eqnarray}\label{chi2}
\chi^{2}=\sum_{i}((I_{data_{i}}-B)+CI_{model_{i}})^{2}
\end{eqnarray}
\noindent where $B$ and $C$ are fitting parameters and $i$ sums
over the pixels in the scattered image. $B$ was calculated once
for each scattered image, and was not adjusted when comparing
different ODF's.  For more details of the model and analysis of
the intraparticle diffraction images, refer to Appendix
\ref{appendix1}.

To measure the orientational distribution function from the
interparticle peak, the method of Oldenbourg et al. was used,
because the scatter consists of only one peak. In this method,
Equation \ref{I_q} simplifies to a one dimensional integral at
constant q$_{r}$. This method is identical to that frequently used
for analyzing thermotropic interparticle scatter, with the
exception that Oldenbourg's method approximates the single rod
scattering, $I_{s}(\Omega)$, as being proportional to
1/$\sin(\omega)$ for small $\theta$. In this equation, $\omega$ is
the angle between the rod and the incident beam. This is in
contrast to most analysis done on interparticle interference in
the past, in which $I_{s}(\Omega)=1$
\cite{Leadbetter78,Davidson95,Deutsch91}. The $1/\sin(\omega)$
proportionality attempts to include finite size of the rod into
the calculation of the ODF. For more details refer to Appendix
\ref{appendix1}.

\section{Results and Discussion \label{results}}
\subsection{Spatial Ordering \label{spatial}} 
\begin{figure}
\epsfig{file=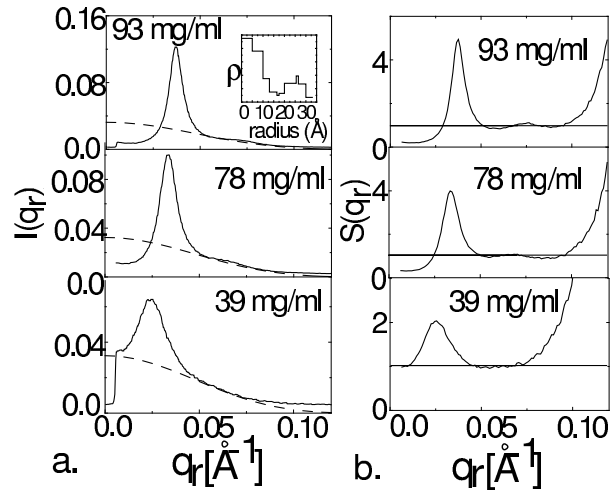, width=8cm}
\caption{\label{eqsfpeaks} Equatorial intensity profile, $I(q_r)$,
and equatorial structure factor, $S(q_r)$, for three
representative samples at 10 mM ionic strength pH 8.2. Smaller
inset graph is the binned radially averaged electron density
$\rho$ used to calculate the equatorial form factor shown as a
dashed line. The deviation of the structure factor from one at
high $q_r$ is due to both background noise in $I(q_r)$, which
hides the actual form factor and a loss of accuracy in the model
form factor at high $q_r$.}
\end{figure}

The location of the
maximum, $q_m$, of the first interference peak and its radial
width, $\Delta q_m$, were measured along the equator, $q_z=0$ in
order to obtain information about the spatial ordering of the
system. Because we are only analyzing data along the equator,
these properties can be determined by dividing the equatorial form
factor, $F(q_r,0)$, from the scattered intensity peaks,
$I(q_r,0)$, and then by fitting the remaining structure factor
peak, $S(q_r,0)$, to a gaussian
$S(q_{r},0)=e^{-(q_{m}-q_{r})^{2}/2(\Delta q_{m})^{2}}$ as done in
Ref. \cite{Schneider80} . $I(q_r,0)$ and $S(q_r,0)$ are shown in
Fig. \ref{eqsfpeaks} for three different samples. The equatorial
form factor scatter was approximated by the Fourier transform of
the known equatorial projection of the cylindrically averaged
electron density of fd \cite{Mcpherson84}. The electron density
was approximated by binning the radial electron density into 10
sections as illustrated in the inset of Fig. \ref{eqsfpeaks}a. The
location of the equatorial peaks produced by the Fourier transform
of the electron density agree with the equatorial form factor data
obtained at higher angle, but the increase in $S(q_{r},0)$ at high
$q_r$ shown in Fig. \ref{eqsfpeaks} indicates that this
approximation is only qualitatively correct at high $q_r$ and that
the presence of background noise in the interparticle diffraction
data hides any high $q_r$ form factor information. At high
concentrations the scattered intensity is much stronger compared
to the readout noise and as a result we are able to analyze the
structure factor data to higher $q_r$ than at low concentrations.

\begin{figure}
\epsfig{file=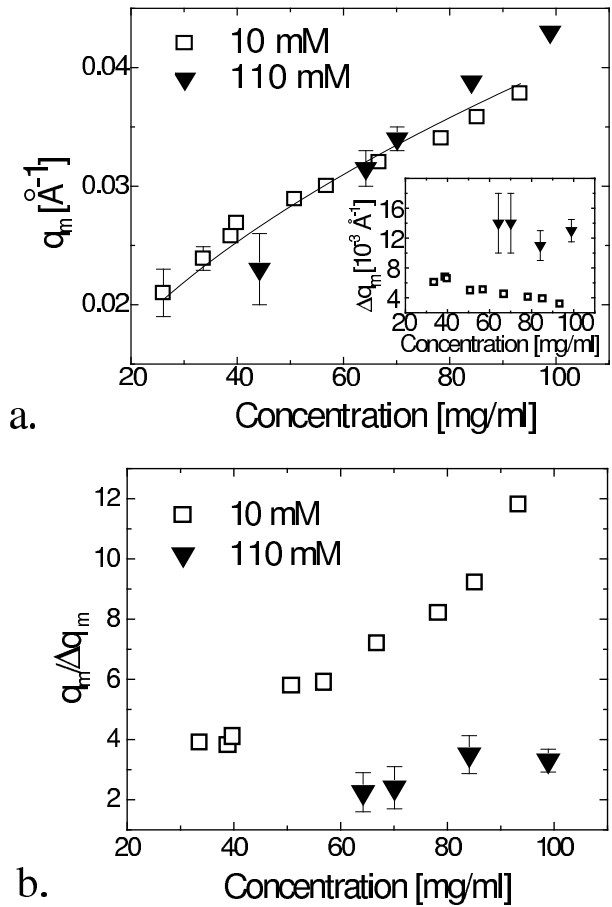,width=8cm} \caption{\label{sfpeakvsc}
(a) The concentration dependance of the maximum $q_m$ of the
interparticle interference peak. The average rod separation is a
distance of 2$\pi/\mbox{q}_{m}$ \AA. The equation of the curve
fitted to the combined data sets is $q_m=0.004c^{1/2}$. The inset
graph shows concentration dependance of the variance of the
interference peaks $\Delta q_m$. (b) The concentration dependance
of $q_m/\Delta q_m$ the number of rods per correlation length.
Squares ($\square$) are at 10 mM and triangles
($\blacktriangledown$) are at 110 mM ionic strength pH 8.2. }
\end{figure}

The $q_m$ and $\Delta q_m$ measured are plotted as a function of
concentration for two different ionic strengths in Fig.
\ref{sfpeakvsc}a. With increasing concentration, the average rod
separation decreases as $c^{-1/2}$ ($q_m \propto c^{1/2}$) as
expected for any rod system \cite{Maier92}. At a given
concentration the rod separation remains constant and the variance
increases with decreasing ionic strength. The electrostatic
repulsion present between the rods causes the rods to maintain the
maximum separation possible, but a smaller effective diameter at
low ionic strength allows for more fluctuations. The number of
rods per correlation length $q_m/\Delta q_m$ is plotted as a
function of concentration in Fig. \ref{sfpeakvsc}b. The
concentration dependance of $q_m/\Delta q_m$is much more
significant at 10 mM ionic strength, than at 110 mM, indicating
that at high ionic strength the rods are less correlated. It is
interesting to note that the second interference peak is much
weaker than the first interference peak, indicating a large
Debye-Waller factor. This is in contrast to charged 3D spherical
and 2D disk systems which show a much stronger second, and even
third interference peak \cite{Asgari01,Rino96}. The structure
factor of nematic fd also contrasts that of nematic end-to-end
aggregated TMV, a very rigid rod, which has a structure factor
closely resembling that of the 2D disk systems \cite{Schneider80}.
One way to interpret the large and sharp first peak in the
structure factor of fd is that flexible nematic rods have long
range spatial correlations similar to a dense fluid of disks.
However the absence of secondary peaks in the structure factor
implies that fd particles have a greater degree of positional
disorder about their average position than do disks. Perhaps the
flexibility of fd accounts for this dramatic difference in spatial
organization.

\subsection{Orientational Ordering \label{angular}}
By examining the $\chi^{2}$ values obtained from the inter- and
intra-particle scatter analysis, and the residues
(I$_{data}$-I$_{fit}$) from the interparticle scatter analysis we
determined that analysis of x-ray diffraction data does not yield
a unique orientational distribution function. The Gaussian and the
Onsager distribution function each fit the intensity data equally
well when comparing residues and $\chi^2$ values from each of the
two functions. However, we were able to eliminate Oldenbourg's
distribution function from the possible ODF forms because it did
not accurately model the tails of the diffraction data at low
concentration. This insensitivity of x-ray diffraction to the
exact form of the ODF was predicted by Hamley who showed that
x-ray patterns are insensitive to higher order terms in the
spherical harmonic expansion of the orientational distribution
function and therefore only an approximation to the full
orientational distribution function can be found \cite{Hamley91}.

\begin{figure}
\epsfig{file=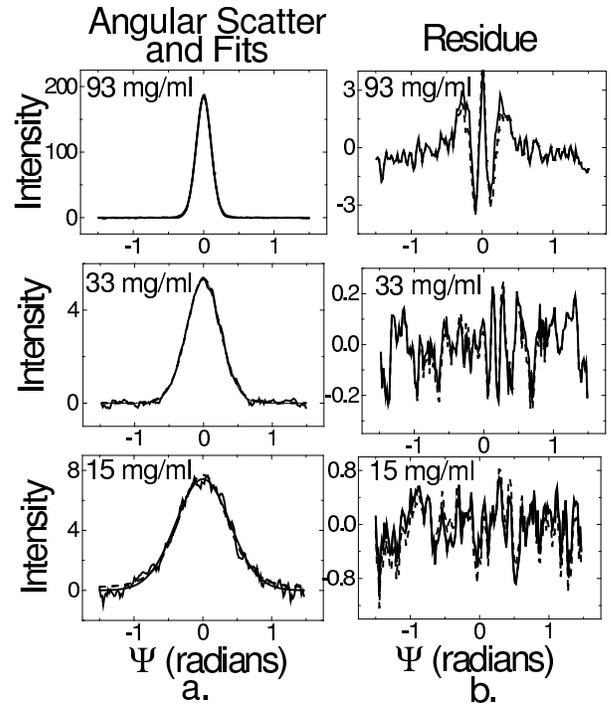,width=8cm} \caption{\label{sfbw.fig}
(a) Angular intensity scan at $q_r$=0.07 $\pm$ 0.001\AA$^{-1}$
from the three data scatter shown in Fig. \ref{sfbw1.fig}a with
best-fit curves calculated from the three trial ODF. Solid fit
line represents fit of both the Gaussian and Onsager ODF's, dotted
line is the fit of Oldenbourg's ODF. (b) Residue ($I_{data}$
-$I_{fit}$) plot. $\Psi$ is illustrated in Fig. \ref{sfbw1.fig}a.}
\end{figure}

\begin{figure}
\epsfig{file=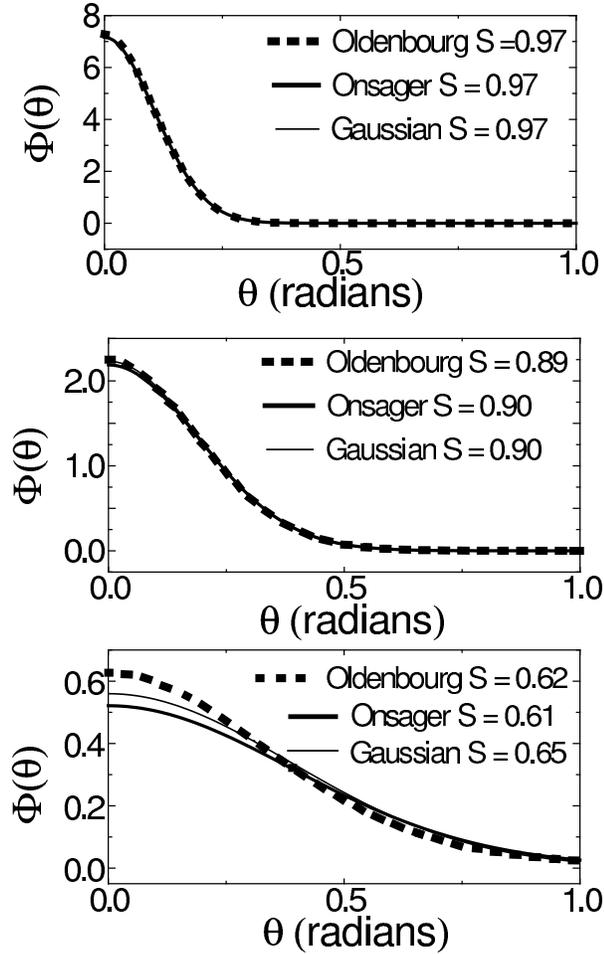,width=8cm} \caption{\label{odf.fig}
Orientational distribution functions calculated from the
interparticle angular intensity scan at constant radius $q_r$=0.07
$\pm$ 0.001\AA$^{-1}$ shown in Fig. \ref{sfbw.fig}b. Gaussian
(thin solid line) , Oldenbourg (dotted line) and Onsager (thick
solid line) ODF are shown. Order parameters shown are calculated
from each ODF. From top to bottom the concentration of the samples
are 93 mg/ml, 33 mg/ml, and 15.5 mg/ml. The ionic strength of the
samples is 10 mM, pH 8.2. Axes of contour plot are labeled in
\AA$^{-1}$.}
\end{figure}

\begin{figure}
\epsfig{file=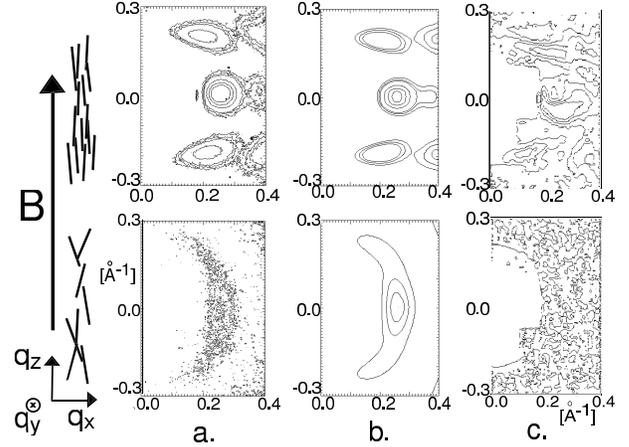,width=8cm} \caption{\label{ffbw.fig}
(a) Contour plots from Fig. \ref{sfbw1.fig}b of scattering from
nematic fd samples due to intraparticle interference. Bottom
scatter is at 15.5 mg/ml and top is at 93 mg/ml. Samples were at
10 mM ionic strength pH 8.2. (b) Simulated intraparticle scatter
using a Gaussian ODF which best fit the intraparticle scatter
shown in (a). (c) Residue ($I_{data}-I_{fit}$)/$I_{fit}$ plot.
Maximum residues in (c) are $\pm$10\%. The axis of the plots are
labeled in \AA$^{-1}$.}
\end{figure}

To demonstrate this assertion, the scattered interparticle
intensity at a constant radius of $q_r$ = 0.07 $\pm$ 0.001
\AA$^{-1}$ is plotted in Fig. \ref{sfbw.fig}a with the best-fit
model intensities for each of the three ODFs. $\Psi$ is the angle
from the equator on the detector film. The actual best-fit
orientational distribution functions calculated from these
interparticle angular scans are shown in Fig. \ref{odf.fig}. The
residues calculated from the interparticle and intraparticle
interference results for the three samples are illustrated in
Figs. \ref{sfbw.fig}b and \ref{ffbw.fig}c, respectively. The
intraparticle scatter residues shown are for the scattered
intensity shown in \ref{ffbw.fig}a minus the model images shown in
Fig. \ref{ffbw.fig}b created with the Gaussian ODF. The
intraparticle model scatter produced relatively uniform residues
indicating that it was a qualitatively good model. In two
dimensions, we were unable to distinguish differences between
residue plots of ODFs of the same width, therefore residue
analysis was limited to the interparticle scatter.
At high concentration small systematic disagreements between the
best-fit models and the data are most visible in the residue plots
in Figs. \ref{sfbw.fig}b and \ref{ffbw.fig}c, but each of the
three models and their respective ODFs are nearly
indistinguishable. Except at low concentration, the best-fit model
intensities obtained from the three distribution functions can not
be distinguished from one another both by analyzing residue plots
and by comparing minimum $\chi^2$ values computed from the fitting
routine. At low concentration the systematic disagreements between
the data and the fits are lost in the noise, but disagreements in
fits from different ODFs become visible. The best fit model
intensities from the Gaussian and Onsager ODFs are
indistinguishable, but the residues from the Oldenbourg ODF show
disagreement, and the fits are systematically higher than the
background scatter at high angle $\Psi$. At the isotropic-nematic
transition the $\chi^2$ values computed from the Oldenbourg ODF
were also consistently higher. The calculated Oldenbourg ODF also
looks significantly different from the calculated Gaussian and
Onsager ODFs. From these qualitative observations we argue that
the distribution function used by Oldenbourg et al. does not
describe our diffraction data as well as the Gaussian and the
Onsager distribution function at low concentrations. The Gaussian
and the Onsager orientational distribution functions fit the
diffraction data equally well.

Because of small differences in the trial orientational
distribution functions (as illustrated in Fig. \ref{odf.fig}),
best-fit ODFs vary slightly in their width, and subsequently
returned slightly different order parameters. But, the order
parameters calculated from the best-fit Gaussian and Onsager ODFs
were in agreement with one another within the experimental
uncertainty of $\Delta S/S<6\%$. Order parameters calculated with
the Oldenbourg ODF were in common agreement at high
concentrations, where model scatter agreed with the data. The
nematic order parameter calculated at multiple $q_{r}$ across
interparticle peak also remained relatively constant, $\Delta
S/S\leq 4\%$. Because we can not distinguish between the Gaussian
and the Onsager model scatter, the order parameters to be
presented henceforth are an average of the values calculated from
only the Gaussian and the Onsager ODF, and the uncertainty on the
values given are a combination of experimental error and
uncertainty due to variation in order parameters from two trial
ODFs.

\begin{figure}  \epsfig{file=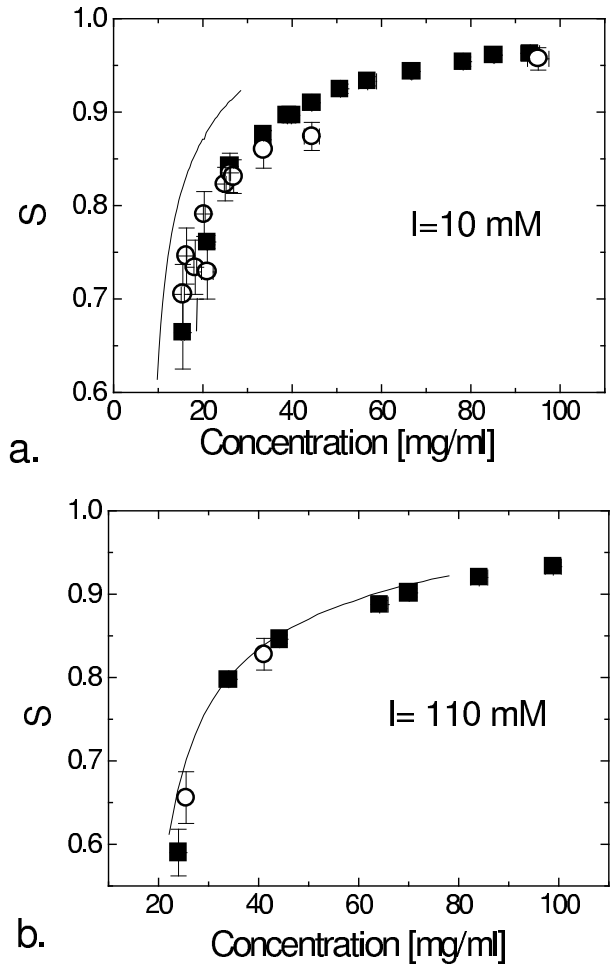,width=8cm}
\caption{\label{SvsC.fig} Concentration dependence of the nematic
order parameter. (a) is at 10 mM, and (b) is at 110 mM ionic
strength and pH 8.2. Squares ($\blacksquare$) are from the
interparticle interference peak, and open circles ($\circ$) are
results from the intraparticle peak. The solid lines shown are for
a scaled particle theory for charged semi-flexible rods.}
\end{figure}

\begin{figure}
\epsfig{file=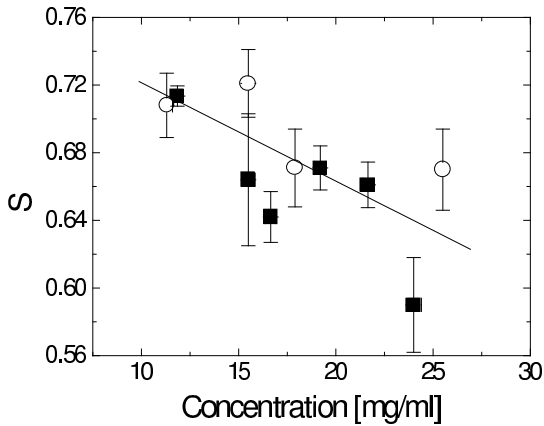,width=8cm}
\caption{\label{compareIN.fig} Concentration dependence of the
order parameter of the nematic phase co-existing with the
isotropic phase as determined from the intraparticle peak (open
circles $\circ$) and the interparticle peak (solid squares
$\blacksquare$). Increasing concentration of the coexistence
concentration is due to increasing ionic strength \cite{Tang95}.
The solid line is a linear fit to the combined sets of data and is
presented as a guide to the eye.}
\end{figure}

The concentration dependance of the order parameters was measured
from both the interparticle and intraparticle peaks and the
resulting values are graphed in Fig. \ref{SvsC.fig}. In Fig.
\ref{compareIN.fig} the order parameter of the nematic phase in
coexistence with the isotropic phase is plotted for five different
ionic strengths as a function of concentration. The coexistence
concentrations are an increasing function of ionic strength. Our
analysis shows that the order parameters calculated from the
interparticle and intraparticle scatter are consistent with one
another both as a function of concentration and of ionic strength,
indicating that correlations in the interparticle peak do not
visibly change measured nematic order parameters.

Fig. \ref{SvsC.fig}a shows data obtained at an ionic strength of
10 mM and Fig. \ref{SvsC.fig}b shows data obtained at an ionic
strength of 110 mM, pH 8.2. With increasing concentration, the
order parameter increases until it saturates near $S=1$, and at
constant concentration, the nematic order parameter decreases with
increasing ionic strength. At low concentrations, the scattered
intensity is spread over a large area due to the broad
orientational distribution function, which leads to a large
decrease in the signal to noise ratio. As a result, the variation
in the calculated order parameters increases, reaching a maximum
of $\Delta S/S\leq$ 10\%. The theoretical curves shown in Fig.
\ref{SvsC.fig} were computed from a scaled-particle theory which
includes semi-flexibility in the orientational entropy and
electrostatic interactions through an effective diameter. This
calculation is outlined in detail in Appendix \ref{appendix2}. The
Onsager ODF was used in calculating these theoretical curves. Our
results qualitatively agree with this theory at low ionic
strength, and quantitatively agree at high ionic strength.


\begin{figure}
\epsfig{file=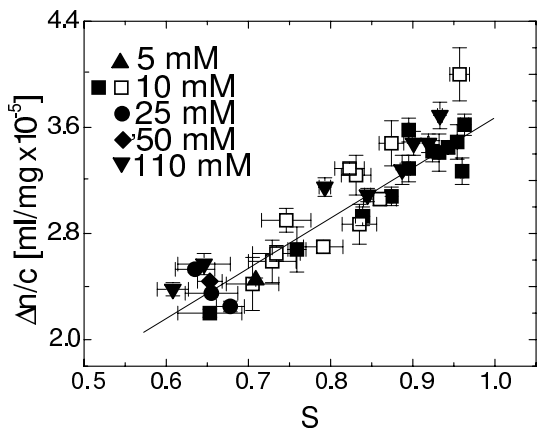,width=8cm} \caption{\label{biref.fig}
Comparison of measured birefringence $\Delta n/c$ to the deduced
x-ray order parameter $S$. Open shapes are from intraparticle
interference peak measurements. Closed shapes are from
interparticle interference measurements. The equation of the
fitted line is $\Delta n/c=(3.8 \pm 0.3)$S$-(0.11\pm 0.19)$ where
$\Delta n/c$ is in units of $10^{-5}$ ml/mg.}
\end{figure}

The values for the order parameter calculated from the x-ray
diffraction data were also compared to birefringence measurements
for the whole range of concentrations and the results are plotted
in Fig. \ref{biref.fig}. Theoretically we expect
\begin{eqnarray}
\frac{\Delta n}{c}=S \frac{\Delta n_{\mbox{\scriptsize{sat}}}}{c},
\end{eqnarray}
\noindent where $\Delta n$ is the sample birefringence and $\Delta
n_{\mbox{\scriptsize{sat}}}$ is the birefringence of perfectly
aligned fd \cite{Maret85}. The saturation birefringence per unit
concentration was measured as $\Delta
n_{\mbox{\scriptsize{sat}}}/c=3.8\times10^{-5} \pm
0.3\times10^{-5}$ ml/mg using data from samples at five different
ionic strengths. Our value for the saturation birefringence is
lower than the previously calculated value by Torbet et. al. of
$\Delta n_{\mbox{\scriptsize{sat}}}/c=6\times10^{-5}$ ml/mg which
was calculated for solutions of fd at 16 mg/ml in 10 mM Tris-HCL
buffer at pH 7.5 placed in a 2-4 T magnetic field \cite{Torbet81}.

\begin{figure}
\epsfig{file=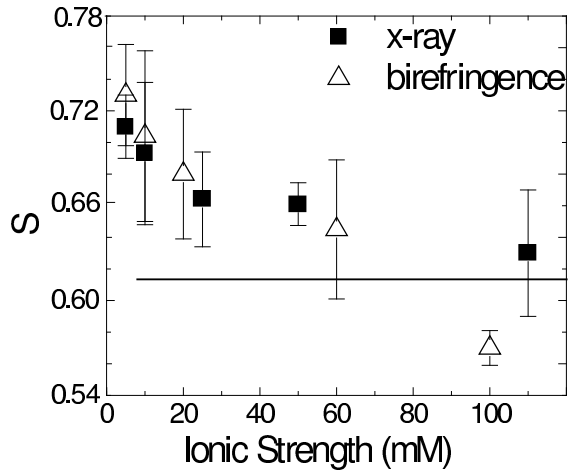,width=8cm}
\caption{\label{SvsISatIN.fig}Ionic strength dependance of the
order parameter of the nematic phase co-existing with the
isotropic phase as calculated by x-ray diffraction measurements
(solid squares $\blacksquare$) and birefringence measurements
(open triangles $\vartriangle$). Solid line shows order parameter
predicted by scaled particle theory for charged semi-flexible rods
as in Fig. \ref{SvsC.fig}.}
\end{figure}

For long rods Onsager predicts that the nematic order parameter at
coexistence remains constant independent of ionic strength, but in
Figs. \ref{compareIN.fig} and \ref{SvsISatIN.fig}, a weak
dependance of the order parameter with ionic strength is seen. In
Fig. \ref{SvsISatIN.fig}, the ionic strength dependance of the
nematic order parameter at coexistence is plotted as deduced from
both x-ray diffraction and birefringence measurements. The change
in ionic strength from 5 mM to 110 mM corresponds to an
$L/D_{\mbox{\scriptsize{eff}}}$ for the rods changing from
$\approx 40$ to $\approx 85$. As the effective aspect ratio for
our rods approaches the long rod limit,
$L/D_{\mbox{\scriptsize{eff}}}> 100$, the coexistence order
parameter decreases, approaching the theoretically predicted value
of $S= 0.55$, as calculated by Chen for long semi-flexible rods
with a length to persistence length ratio, $L/p = 0.4$
\cite{Chen94}. Even though the persistence length of fd virus is
more than twice its contour length, and thus can be considered
fairly rigid, all of our co-existing samples had a nematic order
parameter significantly lower than the Onsager prediction of
$S=0.79$ as measured by both diffraction and birefringence.

Stroobants et al., have shown that there is an additional
electrostatic twisting factor which acts to misalign adjacent
particles and may effect the nematic order parameter at
coexistence \cite{Stroobants86}. This effect scales as
$h=\kappa^{-1}/D_{\mbox{\scriptsize{eff}}}$ where $\kappa^{-1}$ is
the Debye screening length. The effect of $h$ on the coexistence
concentrations of the system is predicted to be small
\cite{Fraden95}, and the predicted decrease in the nematic order
parameter is only 2.4\% when increasing ionic strength from 5mM to
110mM, whereas we measure an decrease in the order parameter of
$\approx$ 15\%. Nevertheless, the observed trend of decreasing
order parameter with increasing ionic strength indicates that the
electrostatic twisting effect might be influencing the system in
addition to the changing effective aspect ratio of the charged
virus. It is also important to note that below about 10 mM, the
concentration of the virus may begin to have an effect on the
ionic strength of the solution in that the concentration of the
virus counterions may act to increase the ionic strength. Overall,
we expect and observe better agreement with the theoretical
predictions for the nematic order parameter at high ionic
strength.

\section{Conclusion \label{concl}}
To summarize, we have observed, as predicted by Hamley, that the
method of using x-ray diffraction to calculate the orientational
distribution function is insensitive to the details of the form of
distribution function used. Nevertheless, we were able to rule out
the function used by Oldenbourg et al. because we could
qualitatively see that models created using this function did not
fit the data at low concentration and at high angle $\Psi$ from
the equator on the detector. The Onsager and Gaussian trial
angular distribution functions fit the angular distribution of
both the intraparticle and interparticle diffraction peaks equally
well and returned similar values for the nematic order parameter.
The concentration dependence of the nematic order parameter at
high ionic strength, or large $L/D_{\mbox{\scriptsize{eff}}}$, as
determined from both the interparticle and intraparticle scatter
agrees with that predicted by a scaled particle theory of charged
semi-flexible rods. At low ionic strength, theoretical predictions
qualitatively reproduce the concentration dependance of the order
parameter. Similar agreement of the concentration dependence of
nematic ordering to Onsager's theory has been measured for the
system of persistence lengthed DNA fragments using neutron
scattering \cite{vandermaarel94,vandermaarel95}. This similarity
demonstrates the universality of Onsager's theory and its
applicability to semi-flexible systems.

The nematic order parameters derived from both interparticle and
intraparticle scatter return similar results, implying that it is
sufficient to use the easier, one dimensional analysis of the
interparticle interference peak to calculate nematic order
parameters as has been done for many years for thermotropic liquid
crystals. It has also been shown that the relationship between the
birefringence and the nematic order parameter as calculated by
x-ray diffraction is linear. From this relationship the saturation
birefringence of fd was calculated. Subsequently, the order
parameter can also be obtained simply by measuring the
birefringence of a sample of nematic fd and rescaling it by the
saturation birefringence. We note that the birefringence
measurements were much less repeatable than diffraction
measurements, as can be observed by the large error bars
throughout the entire range of data shown in Fig. \ref{biref.fig}.

At high ionic strength, or large effective aspect ratio, we
observed that the order parameter of the nematic phase coexisting
with the isotropic phase was $S\approx0.6$, close to the
theoretically predicted value for semi-flexible rods and
significantly lower than the theoretical value of $S=0.79$ for
rigid rods. With decreasing ionic strength however, a weak
systematic increase in the nematic coexistence order parameter was
found. This is consistent with both an decrease in the twist
parameter and a deviation of $L/D_{\mbox{\scriptsize{eff}}}$ from
the long rod limit. In order to fully understand the interactions
which are producing the nematic phase diagrams, particularly at
lower ionic strength where $L/D_{\mbox{\scriptsize{eff}}}$ is
small new theories and simulations need to be developed which
include a more complete picture of the complicated electrostatic
interactions.

\begin{acknowledgments}
We thank E. Belamie for helping with data collection at the
beamline. Work at Brandeis is supported by the NSF(DMR-0088008).
Work at Yale was supported by the NSF(DMR-0071755). 8-ID beamline
at the Advanced Photon Source is supported by the DOE
(DE-FG02-96ER45593) and NSERC. The Advanced Photon Source is
supported by the DOE (W-31-109-Eng-38).
\end{acknowledgments}

\appendix
\section{X-ray Diffraction Angular Analysis\label{appendix1}}

If we assume we are at sufficiently high scattering angle where
intensity variations due to interparticle interactions are
negligible then $S(q_{r},q_z)=1$, and we can measure the
orientational distribution function from intraparticle
interference by comparing it to a simulated scatter created from
the evaluation of Eq. \ref{I_q}. To evaluate Eq. \ref{I_q} a three
dimensional model for the single rod form factor ${I_s(q_r,q_z)}$
was developed.

\begin{figure*}
\centerline{\epsfig{file=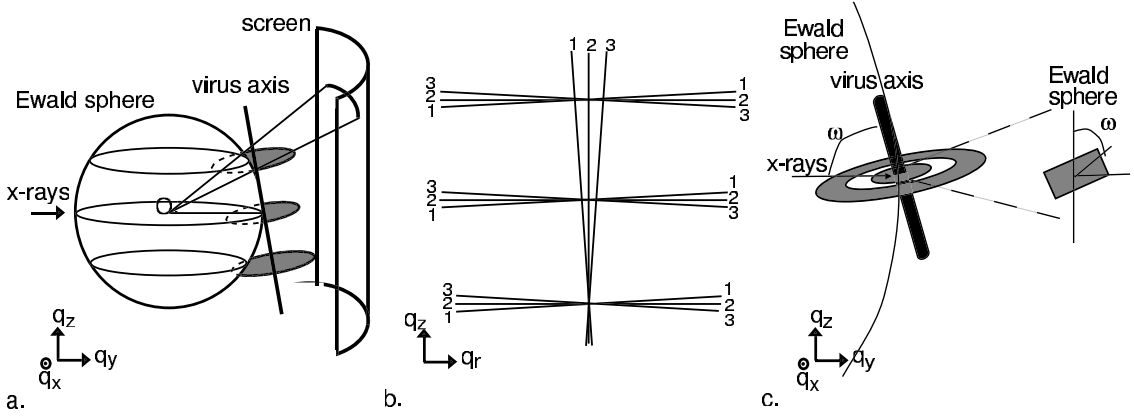, width=16cm}}
\caption[]{\label{threepicsforanalysis}(a)Schematic of the Fourier
space of a single rod tilted at a slight angle. (b) Schematic
showing how the intensity along the center of the layer lines
decreases as $q_r$ when there is a small amount of angular
disorder. Three rod axes (vertical) are labeled 1,2,3 along with
their corresponding contribution to layer lines 0,1,2 (horizontal)
as shown. (c) Schematic showing effect of the thickness of the
form factor disks on the scatter with changing $\omega$. The right
hand image in (c) is an enlargement of the equatorial intersection
of the Ewald sphere and $I_s$.}
\end{figure*}

A long rod Fourier transforms as a disk of thickness $2\pi/L$
oriented perpendicular to the long axis of the rod. Because of the
helical periodic structure along the long axis of fd, the Fourier
transform of a single fd consists of a series of disks separated
by a distance proportional to the reciprocal of the period
\cite{Holmesbk}. This is shown schematically in Fig.
\ref{threepicsforanalysis}a. The radial intensity along these
disks is a summation of Bessel functions whose exact form depends
on the structure of the rod. When projected onto a screen these
disks are visible as layer lines. The images shown in Figure
\ref{ffbw.fig}b show the zeroth and $\pm$ first layer lines. For
our model, the radial intensities of the disks were approximated
by the scattered intensities along the middle of the zeroth and
$\pm$ first layer lines, $I_{m}$, of our most aligned nematic
sample, times the radius $q_{r}$ at which that intensity is
located and the width, $\alpha$, of the Gaussian ODF,
\begin{eqnarray}
I_{s}=I_{m}\sqrt{2\pi}\alpha q_{r}. \
\end{eqnarray}
\noindent $\sqrt{2\pi}\alpha q_r$ is the disorientation correction
term. For a small amount of disorientation of rods, the radial
intensity decreases as $1/q_r$. The effect of the disorientation
is illustrated in Fig. \ref{threepicsforanalysis}b. This
approximation method was developed by Holmes and Leigh, and is
valid if the sample from which the $I_m$ is taken was well aligned
\cite{Holmes74}. The nematic order parameter of our most aligned
sample was $S=0.96$ as measured from the interparticle
interference peak.

In order to model diffraction from a nematic phase of fd, this
single particle scattered intensity is multiplied by a test ODF
and integrated over all possible angles of orientation, as in Eq.
\ref{I_q}. The intersection of the resulting three dimensional
nematic form factor and the Ewald sphere is then ``projected''
onto a two dimensional ``screen'' and a final two dimensional
image is created, as shown in Fig. \ref{threepicsforanalysis}a.
The width of the Ewald sphere was chosen to equal the energy
uncertainty of the experimental beam $\Delta E/E= 5\pm 1\times
10^{-4}$. The ``screen'' image is then convolved with the Gaussian
point spread function of the experimental x-ray beam approximated
as exp[$-r^{2}/2\sigma^{2}]$ with $\sigma = 0.0063$ \AA. A series
of two dimensional images were made for different orientational
distribution functions with different amounts of disorder,
examples can be seen in Fig. \ref{ffbw.fig}b. All collected data
within the range of $q_r=0.19-0.33$\AA is fitted to these
simulated images to find the ODF which minimized a computed
chi-squared value shown in Eq. \ref{chi2}. This $q$-range
encompassed the innermost peak on each of the three layer lines
visible in the intraparticle interference scatter.

To measure the orientational distribution function from the
interparticle peak, the method of Oldenbourg et al. was used.
Because we measure the angular spread of only one diffraction
peak, Eq. \ref{I_q} simplifies to a one dimensional integral at
constant $q_{r}$:
\begin{equation}
\label{I_q_1D} I({\Psi}) =\int {\Phi}({\theta}){I_{s}({\omega})}
\sin{\omega} d {\omega}
\end{equation}
\noindent where $I({\Psi})$ is the angular intensity distribution
along an arc drawn at constant radius, $\Psi$ is measured from the
equator on the detector film, $\Phi(\theta)$ is the angular
distribution function of the rods and $\omega$ is the angle
between the rod and the incoming beam. $\Psi$, $\theta$ and
$\omega$ are related by $\cos\theta = \cos\Psi \sin\omega$. Even
thought it was originally used for analyzing intraparticle
scatter, this equation is identical to that used for analyzing
thermotropic interparticle scatter, except that Oldenbourg's
method includes a term which accounts for the length of the rod by
defining the single rod scattering as $I_{s}(\omega)=
1/\sin\omega$, for small $\theta$, where $\omega$ is the angle
between the rod axis and the x-ray beam as illustrated in Fig.
\ref{threepicsforanalysis}c. This $1/\sin\omega$ proportionality
comes from the understanding that the Fourier transform of a rod
of finite length is a ring with a finite thickness, and as
$\omega$ decreases $1/\sin\omega$ increases and more of the disk
intersects the Ewald sphere and is subsequently projected onto the
detector screen.

Analysis done on interparticle interference from thermotropic
liquid crystals typically defines $I_{s}(\omega)=1$
\cite{Leadbetter78,Davidson95,Deutsch91}. It has been previously
shown through calculations that neglecting the angular width when
calculating the order parameter from interparticle interference
scatter results in inaccurate values for the nematic order
parameter for $S>0.8$ \cite{Leadbetter78}. But, in our analysis we
observed that changing $I_{s}$ from $1/\sin\omega$ to one in the
interparticle interference scatter analysis did not have a
significant effect on the calculated value of the nematic order
parameter, nor did the $\chi^{2}$ values reveal any information as
to which $I_s$ better describes the data. We chose to include the
effect of rod length in our interparticle scatter analysis to be
consistent with our intraparticle scatter analysis, which requires
a knowledge of the rod length.
\section{Scaled Particle Theory\label{appendix2}}
To compare the experimental results for the order parameter to the
theory, we use the scaled particle expression for free energy of
hard rods as was developed by Cotter and
coworkers~\cite{Cotter78,Cotter79}. The main advantage of the
scaled particle theory is that it takes into account third and all
higher virial coefficients in an approximate way and leads to very
good agreement with simulation results for the I-N
coexistence~\cite{Kramer98}. Therefore this theory should be more
adequate at describing data at higher concentrations of rods. We
also note that the expression for the free energy
(Eq.~\ref{scaled_free_energy}) reduces to Onsager's second virial
approximation for very long rods ($L/D \rightarrow \infty$). The
free energy derived by Cotter is :
\begin{eqnarray}
\label{scaled_free_energy}
    \frac{F(\delta,\phi,\alpha)}{Nk_bT}=\ln(\phi)+\ln(1-\phi)+\sigma(\Phi(\alpha))\nonumber\\
    +\Pi_2(\delta,\alpha) \frac{\phi}{1-\phi}+\frac{1}{2} \Pi_3(\delta,\alpha)
    \left(\frac{\phi}{1-\phi}\right)^2
\end{eqnarray}
\noindent where $\phi$ is the volume fraction of rods
\begin{eqnarray}
    \phi=\frac{N_{rods}}{V}
    \left(\frac{\pi}{6}D^3+\frac{\pi}{4}D^2L
    \right).
\end{eqnarray}
\noindent The coefficients $\Pi_2$ and $\Pi_3$ are given by the
following expressions:
\begin{eqnarray}
\Pi_2(\delta,\alpha)=3+
\frac{3(\delta-1)^2}{3\delta-1)}\xi(\Phi(\alpha)),
\end{eqnarray}
\begin{eqnarray}
\Pi_3(\delta,\alpha)=\frac{12\delta(2\delta-1)}{(3\delta-1)^2}+\frac{12\delta(\delta-1)^2}{(3\delta-1)^2}\xi(\Phi(\alpha))
\end{eqnarray}
\noindent and parameter $\delta$ is the overall length over
diameter ratio of the spherocylinder given by $\delta=(L+D)/D$.
The functions $\xi(\alpha)$ is the excluded volume interaction
between two rods as derived by Onsager
\begin{eqnarray}
\xi (\alpha) =  \frac{2I_2(\alpha)}{\sinh^2(\alpha)}.
\end{eqnarray}
\noindent The expression that accounts for the rotational entropy
of the rods and the entropy associated with the loss of
configurations due to confinement of the bending modes of the
semi-flexible rods in the nematic phase has been derived by
extrapolating between the hard rod and the flexible chain
limits~\cite{Hentschke90,Odijk86,DuPre91}. In this paper the
expression obtained by Hentschke is used for numerical
calculations
\begin{eqnarray}
\sigma(\alpha,\frac{L}{p})=\ln(\alpha) -1 +\pi
e^{-\alpha}+\frac{L}{6p}(\alpha-1)\nonumber\\
+\frac{5}{12}\ln\left(\cosh\left(\frac{L}{p}\frac{\alpha-1}{5}\right)\right)
\end{eqnarray}

After the expression for the free energy is obtained, we use
Onsager approximation for the orientational distribution function
$\Phi(\alpha)$ (Eq. \ref{onsODF}) and minimize the scaled particle
free energy in Eq. ~\ref{scaled_free_energy} with respect to the
parameter $\alpha$ to find the order parameter of the nematic
phase at different rod concentrations. To find out the
concentrations of rods in the coexisting isotropic and nematic
phases we solve the conditions for the equality of the osmotic
pressure and chemical potential.

To take into account the fact the rods are charged, instead of
using the hard core diameter $D$ in our calculations we use an
effective diameter $D_{\mbox{\scriptsize{eff}}}$~\cite{Tang95}.
Strictly speaking this re-scaling procedure by
$D_{\mbox{\scriptsize{eff}}}$ is valid only for densities at which
the system is described by the second virial approximation,
therefore our theoretical prediction has an uncontrolled
approximation. Despite this fact the agreement between the theory
and the experiments is quite satisfactory. It is worth mentioning
that there have been recent effort to extend the validity of the
scaled particle theory to include repulsive interactions, however
this theory was not included in our calculations
~\cite{Kramer99,Kramer00}.


\end{document}